\documentclass[aps,prapplied,reprint,groupedaddress,twocolumn]{revtex4-1}
\usepackage{graphicx,subfigure}
\graphicspath{{figures/}} 
\usepackage{amsmath,amsfonts,amssymb,amsbsy,amscd,amsgen,amsthm,dsfont}
\usepackage[]{natbib}
\usepackage{color}
\usepackage[colorlinks,linkcolor=blue,citecolor = blue,urlcolor  = blue]{hyperref}
\usepackage{comment}
\usepackage{multirow}
\usepackage{bm}
\theoremstyle{definition}

\newcommand{\id}{\mathrm d}
\newcommand{\vc}{\mathbf}

\newcommand{\re}{\mbox{Re}}

\renewcommand{\tilde}{\widetilde}

\begin{document}
\title{Closed-loop adaptive control of extreme events in a turbulent flow}
\author{Mohammad Farazmand}
\thanks{Corresponding author: farazmand@ncsu.edu}
\affiliation{Department of Mathematics, North Carolina State University, 2311 Stinson Dr., Raleigh, NC 27695-8205, USA}
\author{Themistoklis P. Sapsis}
\affiliation{Department of Mechanical Engineering, Massachusetts Institute of Technology, 77 Massachusetts Ave., Cambridge, MA 02139-4307, USA}
\date{\today}
\begin{abstract}
Extreme events that arise spontaneously in chaotic dynamical systems 
often have an adverse impact on the system or the surrounding environment. As such, their mitigation is highly desirable. Here, we introduce a novel control strategy for mitigating extreme events in a turbulent shear flow. The controller combines a probabilistic prediction of the extreme events with a deterministic actuator. The predictions are used to actuate the controller only when an extreme event is imminent. When actuated, the controller only acts on the degrees of freedom that are involved in the formation of the extreme events, exerting minimal interference with the flow dynamics. As a result, the attractors of the controlled and uncontrolled systems share the same chaotic core (containing the non-extreme events) and only differ in the tail of their distributions. We propose that such adaptive low-dimensional controllers should be used to mitigate extreme events in general chaotic dynamical systems, beyond the shear flow considered here.
\end{abstract}
\maketitle

\section{Introduction}
Many chaotic dynamical systems exhibit spontaneous extreme events 
which cause abrupt changes in the state of the system~\cite{scheffer2009,lucarini2016,farazmand2019a}. Well-known 
examples include extreme weather patterns, oceanic rogue waves, earthquakes and shocks in power grids.
Since extreme events cause adverse humanitarian, environmental and financial impacts, their 
mitigation is of great interest.

In order to design control strategies that mitigate the extreme events, it is 
crucial to understand the mechanisms that generate them.
The controller should either disrupt these mechanisms or 
counteract their effects.

Recent studies show that, in many systems, only a few degrees of freedom 
contribute to the formation of extreme events, even though the system as a 
whole may be very high dimensional~\cite{PRE2016,mohamad16,Farazmande1701533,babaee17,Dematteis2018,Majda2019b}.
This raises the prospect of designing simple
low-dimensional controllers that mitigate the extreme events
by only acting on these few degrees of freedom. 
%In high-dimensional systems,
%the extreme event mechanisms are usually non-trivial 
%and their discovery requires a careful analysis of the governing equations of the system. 
%Only recently, a systematic framework 
%was developed to analyze the causal effects that lead to the formation 
%of extreme events~\cite{mohamad16,Farazmande1701533,Majda2019}. This variational framework
%also facilitates the data-driven prediction of extreme events through 
%indicators that give early warning signs of upcoming extreme events.

Here, we explore the feasibility of such simple controllers. 
We require the control design to
have two specific features: 
i) Low dimensionality: The controller should only act
on those degrees of freedom that are involved in the formation of extreme events. 
This allows for the simplest possible control design, and therefore facilitates its practical 
implementation.
ii) Adaptivity: We require the control to automatically actuate only 
when there is a high probability of an imminent extreme event. In other words, the 
control is inactive most of the time. Shortly before an extreme event takes place, 
it becomes active, mitigating the event. The control becomes inactive
again after the extreme event episode. 
This requires a real-time prediction scheme for the extreme events.
This adaptive property is similar to the occasional proportional feedback control~\cite{Hunt1991} 
used to stabilize equilibria and periodic orbits, except that we use a Bayesian probabilistic prediction for 
the occurrence of extreme events. 
\begin{figure}
\centering
\includegraphics[width=.25\textwidth]{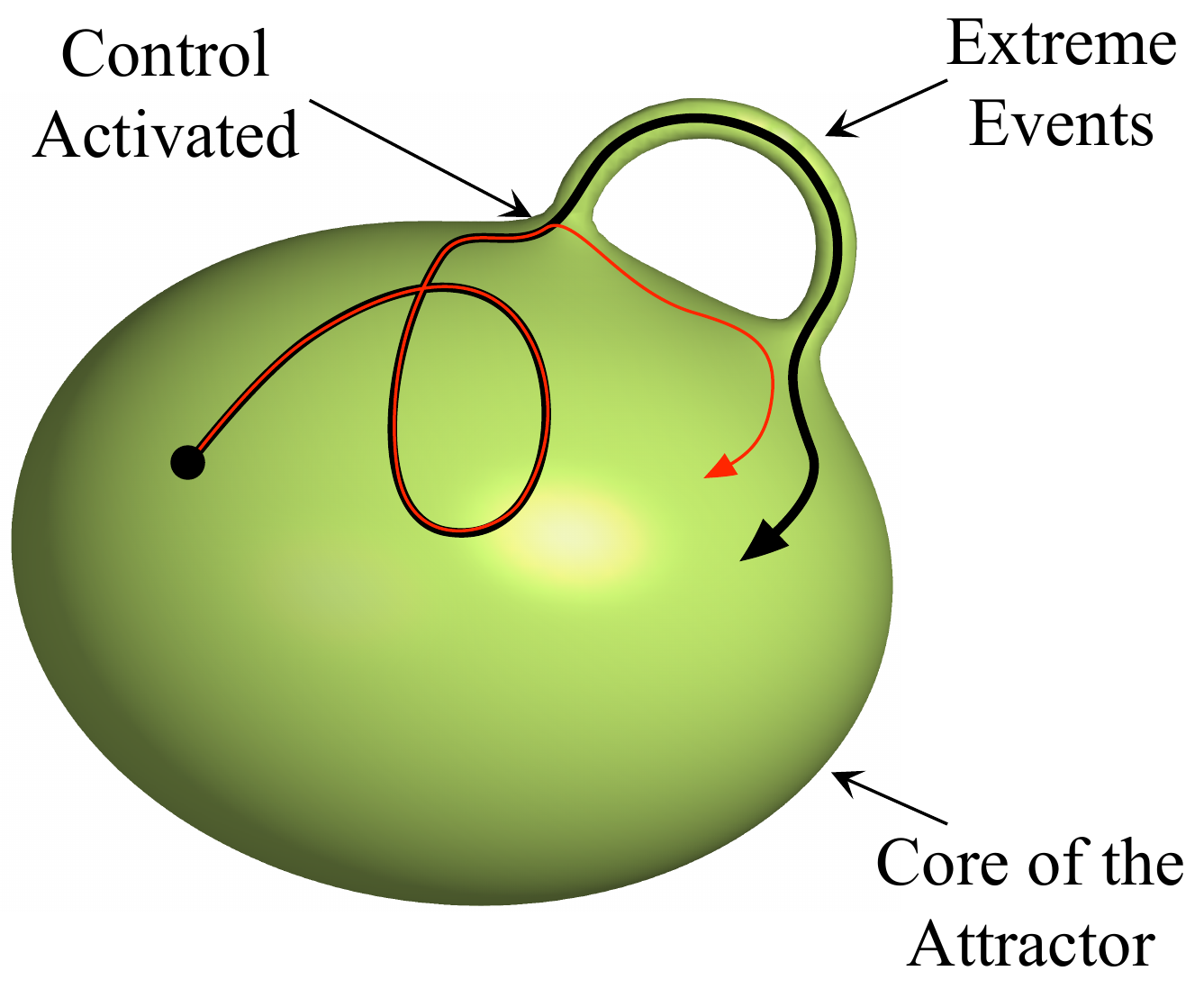}
\caption{Schematic depiction of the system attractor. The thick black curve shows a
trajectory of the uncontrolled system undergoing an extreme event. The thin
red curve marks the trajectory of the controlled system which evades the extreme event. 
The black dot marks the initial state of the system.}
\label{fig:schem_attractor}
\end{figure}

These requirements distinguish our approach from classical control strategies
that seek to suppress the chaotic behavior of the system altogether by 
stabilizing a particular equilibrium state or a periodic 
orbit~\cite{ott1990,pyragas1992,grigoriev1997,Parekh1998,brunton2015,noack2017,Blanchard2019}.
Instead, our approach leaves the chaotic core of the attractor (corresponding to the non-extreme events) intact 
and only prunes the small portion of the attractor that corresponds to extreme events 
(see figure~\ref{fig:schem_attractor}, for an illustration).

Here we demonstrate the feasibility of such extreme event mitigations on a canonical 
turbulent flow: the two-dimensional Navier--Stokes equation
driven by a sinusoidal body force, usually referred to as the Kolmogorov flow~\cite{Arnold1960,obukhov1983,PlSiFi91}.
Extreme events are a common feature of 
moderate and high Reynolds number fluid flow regardless of 
the external forcing or boundary conditions~\cite{aubry88,Meneveau2005,schumacher2014,yeung2015,schmidt2019,blonigan2019,raman2019}. 
These extreme events
can be divided into two broad categories: local and global. Local
extreme events correspond to unusually high velocity gradients 
in a subset of the fluid domain~\cite{schumacher2014,schmidt2019}. In contrast, global extreme events
cannot be pinpointed to a localized event; instead they correspond to 
the space-averaged quantities of the flow~\cite{yeung2015,blonigan2019}. 

The extreme events in Kolmogorov flow
are of the global type and appear as intermittent bursts of the total energy dissipation rate. 
Controlling local extreme events in turbulence requires a predictive scheme that,  in real time,
tracks the location of the extremes in the fluid domain. 
As such, mitigation of the local extreme events seems out of reach at the moment.

\section{Problem set-up}
We consider the incompressible Navier--Stokes equations on the two-dimensional
domain $\Omega=[0,2\pi]\times[0,2\pi]$ with periodic boundary conditions.
Our control strategy is best described in the Fourier space. 
We denote the components of the velocity field by $u_i(\vc x,t)$ ($i=1,2$)
and their Fourier transforms by 
$\hat u_i(\vc k,t)=\int_\Omega u_i(\vc x,t)\exp(-\hat i\vc k\cdot\vc x)\id^2\vc x/(2\pi)^2$
where $\vc k=(k_1,k_2)\in\mathbb Z^2$, $\vc x=(x_1,x_2)\in\Omega$ and $\hat i =\sqrt{-1}$.
The Navier--Stokes equation in the Fourier space reads~\cite{kraichnan1967}
\begin{align}
\partial_t \hat u_i(\vc k,t)= &-\hat i k_mP_{ij}(\vc k)\sum_{\substack{\vc p,\vc q\in\mathbb Z^2 \\ \vc p+\vc q=\vc k}} \hat u_m(\vc p,t)\hat u_j(\vc q,t)\nonumber\\
& -\nu |\vc k|^2\hat u_i(\vc k,t) +\hat f_i(\vc k) + \hat\xi_i(\vc k,t),
\label{eq:nse}
\end{align}
where summation over repeated indices is implied. 
Here, $P_{ij}(\vc k)=\delta_{ij} - k_ik_j/|\vc k|^2$ denotes the Leray projection onto 
divergence-free vector fields where $\delta_{ij}$ is the 
Kronecker delta function. The dimensionless parameter $\nu=\re^{-1}$ is 
the inverse of the Reynolds number $\re$.
The external forcing $\vc f(\vc x)=(\sin(k_f x_2),0)$
is a time-independent shearing body force with the forcing wavenumber $k_f=4$. 
The term $\hat\xi_i(\vc k,t)$ denotes the control
to be discussed shortly.

To solve system~\eqref{eq:nse} numerically, we use a standard pseudo-spectral method with
$2/3$ dealiasing~\cite{Fox1973}.
At the lowest Reynolds number considered here ($\re=40$), we use
$128\times 128$ Fourier modes, while at higher Reynolds numbers we 
use $256\times 256$ modes. 
For the temporal integration, we use the adaptive Runge-Kutta
scheme RK5(4) of Dormand and Prince~\cite{ode45} with
relative and absolute tolerances set to $10^{-5}$.

\section{Uncontrolled System}
\begin{figure}
\centering
\includegraphics[width=.49\textwidth]{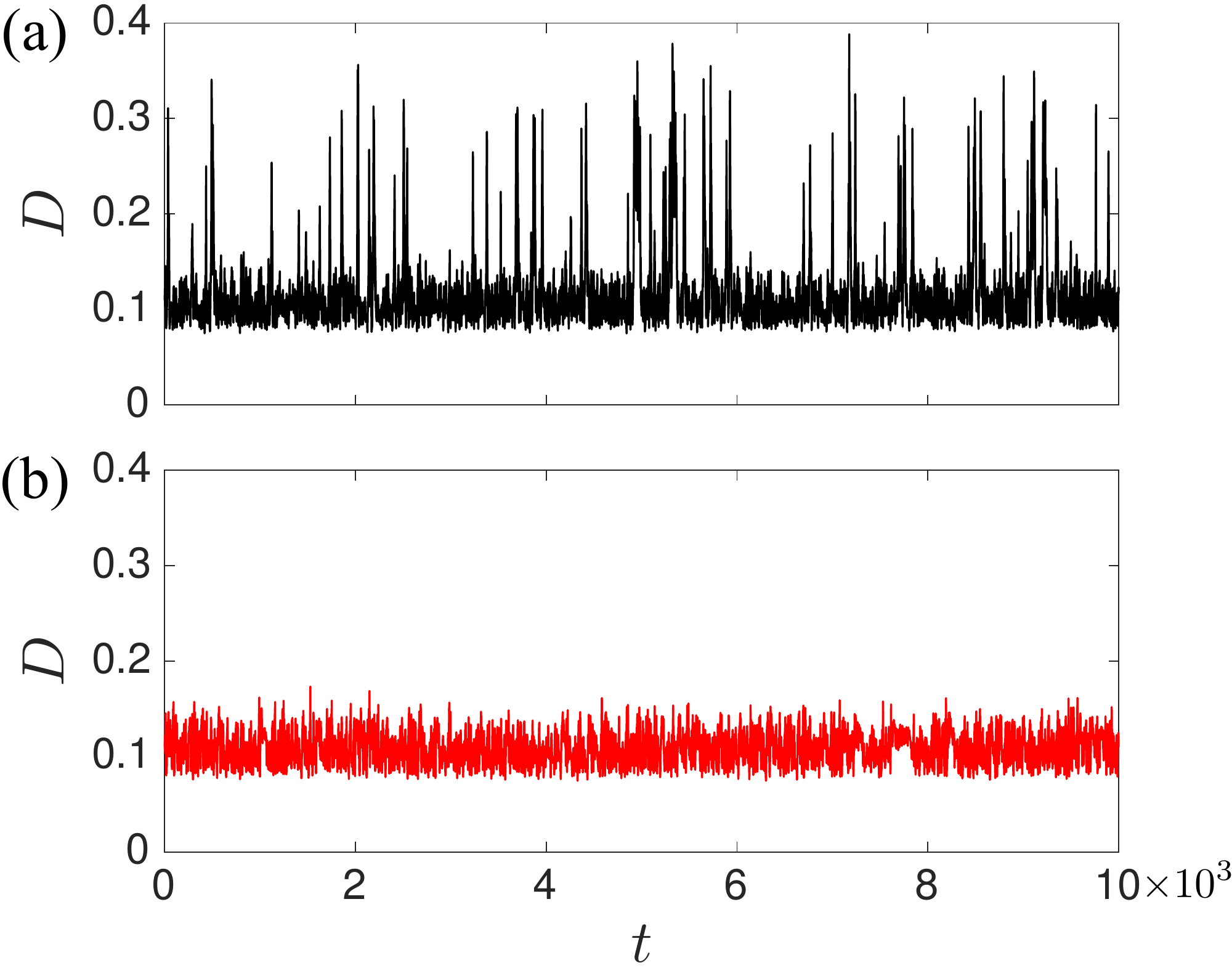}
\caption{The time series of the energy dissipation rate $D$ at
Reynolds number $\re=40$. (a) Uncontrolled system, (b) Controlled system. }
\label{fig:D_ts}
\end{figure}

Much is known about the uncontrolled system 
where $\hat\xi_i\equiv 0$. 
In particular, at Reynold numbers $\re>35$, the uncontrolled system is chaotic with sporadic bursts of the energy 
dissipation rate~\cite{faraz_adjoint}, 
\begin{equation}
D = \frac{\nu}{(2\pi)^2}\int_\Omega |\nabla \vc u|^2\id^2\vc x=
\nu \sum_{\vc k\in\mathbb Z^2} |\vc k|^2|\hat{\vc u}(\vc k)|^2.
\label{eq:Ddef}
\end{equation}
During these bursts, the energy 
dissipation $D$ increases to several standard deviations above its expected value (see figure~\ref{fig:D_ts}(a)).
Using a variational method, 
Farazmand and Sapsis~\cite{Farazmande1701533} showed that these bursts are preceded by 
a nonlinear energy transfer from the Fourier mode $\hat{\vc u}(1,0)$
to the mode $\hat{\vc u}(0,4)$. 

Shortly before an extreme energy dissipation event is 
observed, most of the energy content of the Fourier mode $\hat{\vc u}(1,0)$
is transferred to the Fourier mode $\hat{\vc u}(0,4)$. 
This transfer of energy from the lower mode $(1,0)$ to the higher mode $(0,4)$
leads to an increase in the energy dissipation rate $D$. 
Examining the last identity in equation~\eqref{eq:Ddef} reveals why
such an energy transfer leads to an increase in the energy dissipation.
Since higher Fourier modes are weighted by a larger prefactor $|\vc k|^2$,
transfer of energy from any lower Fourier mode to 
a higher Fourier mode leads to an increase in the energy 
dissipation rate $D$ (assuming that the total kinetic energy does not change significantly). 
In principle, any such downscale transfer of energy
will increase the energy dissipation rate. In Kolmogorov flow, however, 
it is the particular abrupt transfers from the lower Fourier mode 
$\hat{\vc u}(1,0)$ to the higher Fourier mode $\hat{\vc u}(0,4)$
that is responsible for the bursting behavior of the energy dissipation rate~\cite{Farazmande1701533}.

One can go one step further and ask: what triggers this abrupt transfer of energy? Unfortunately, 
the answer to this question is still unknown. Recent results on 
Burgers equation~\cite{Biferale16, Buzzicotti2016}
suggest that the answer lies in resonances between the phases of the
Fourier modes involved in the energy transfer. However, for Kolmogorov flow, 
this possibility remains to 
be investigated. Nonetheless, we show that even the available partial knowledge  
about the precursors to extreme events is sufficient for their prediction
and suppression in the Kolmogorov flow.

\begin{figure*}
\centering
\includegraphics[width=.8\textwidth]{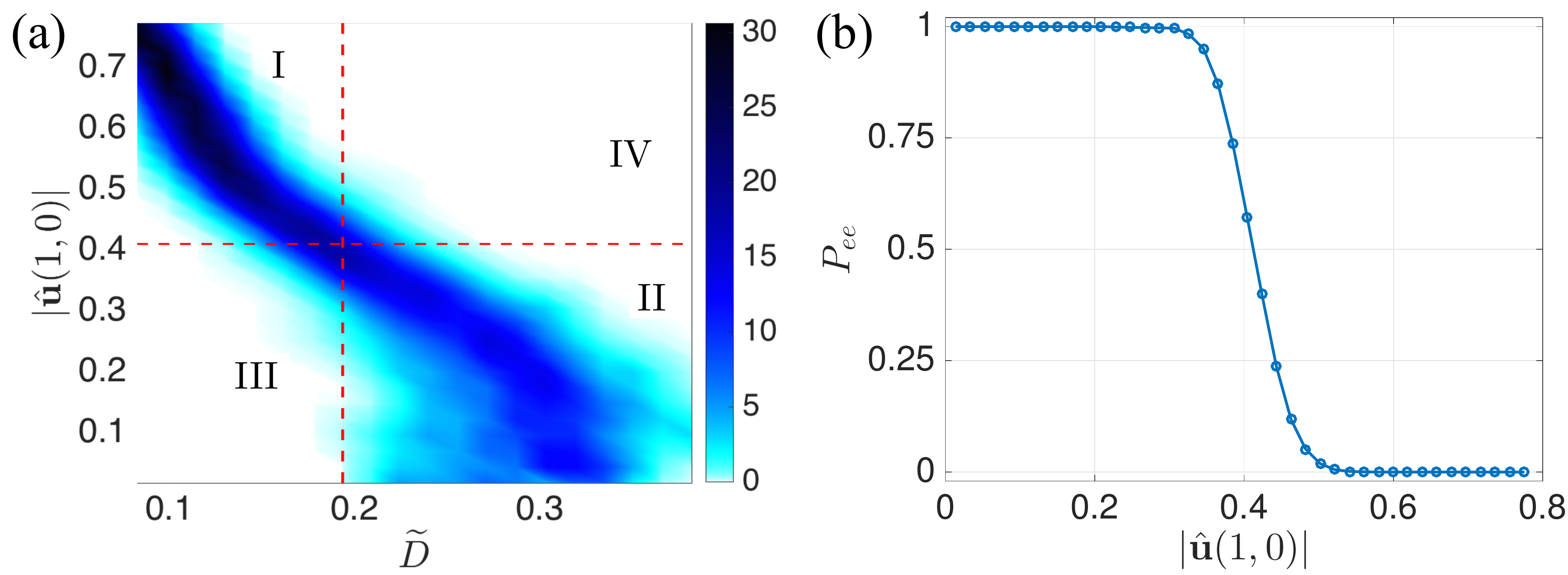}
\caption{Prediction of extreme events at $\re=40$. 
(a) Conditional PDF of
$\widetilde D(t)=\max_{s\in[t+\tau_p,t+\tau_p+\Delta\tau_p]}D(s)$
given $\hat{\vc u}(1,0,t)$. 
(b) Probability $P_{ee}$ of an extreme dissipation event
occurring over the future time interval $[t+\tau_p,t+\tau_p+\Delta\tau_p]$
given the current value of $|\hat{\vc u}(1,0,t)|$.
}
\label{fig:R40_condPDF}
\end{figure*}

In particular, the energy content of 
the mode $\hat{\vc u}(1,0)$ can be used as a predictive indicator for upcoming extremes of the
energy dissipation rate $D$.
To quantify the predictive skill of this indicator, 
we use the conditional probability of 
$\tilde{D}(t)$ given $\lambda(t)=|\hat{\vc u}(1,0,t)|$ at a given time $t$,
where $\tilde D(t)=\max_{s\in[t+\tau_p,t+\tau_p+\Delta\tau_p]}D(s)$
is the maximum value of the energy dissipation rate $D$ over the short future time interval $[t+\tau_p,t+\tau_p+\Delta\tau_p]$.
The prediction time $\tau_p$ determines how far in advance the 
extreme events are predicted. Here, we set $\tau_p=\Delta\tau_p=1.0\simeq 2\tau_e$
which is approximately equal to two eddy turnover times, $\tau_e = \sqrt{\nu/\mathbb E[D]}$.
Here, $\mathbb E$ denotes the expected value.

Figure~\ref{fig:R40_condPDF}(a) shows the conditional probability density 
$p_{\tilde D|\lambda}=p_{\tilde D,\lambda}/p_\lambda$
where $p_\lambda$ is the probability density of $\lambda$, and
$p_{\tilde D,\lambda}$ is the join probability density of $\tilde D$ and $\lambda=|\hat{\vc u}(1,0)|$. This conditional PDF is 
estimated from longterm direct numerical simulations.

The vertical dashed line in figure~\ref{fig:R40_condPDF}(a) marks the threshold $D_e$ for extreme dissipation 
events such that $D>D_e$ constitutes an extreme event. Here the threshold is set to the mean plus two standard deviations of
the dissipation, i.e., $D_e=\mathbb E[D]+2\sigma(D)\simeq 0.2$. The horizontal dashed line
marks the corresponding threshold $\lambda_e$ for the indicator $\lambda=|\hat{\vc u}(1,0)|$.
These two lines divide the conditional PDF plot into four quadrants I-IV as maked on figure~\ref{fig:R40_condPDF}(a).
Below, we describe the significance of each quadrant.

Quadrant I (correct rejections): 
Most of the density of the conditional probability $p_{\tilde D|\lambda}$ is 
concentrated in this quadrant where $|\hat{\vc u}(1,0)|>\lambda_e$ and $\tilde D<D_e$.
The relatively large values of $|\hat{\vc u}(1,0)|$ indicate that no significant nonlinear transfer of energy
from mode $\hat{\vc u}(1,0)$ to mode $\hat{\vc u}(0,4)$ has taken place and 
therefore no upcoming extreme events are expected. 
Since, in this quadrant, we also have $\tilde D<D_e$, this implies that the indicator 
correctly predicted no upcoming extreme events.

Quadrant II (correct predictions):
In this quadrant we have $|\hat{\vc u}(1,0)|<\lambda_e$ and $\tilde D>D_e$.
As mentioned earlier, prior to an extreme event,
$|\hat{\vc u}(1,0)|$ becomes small since most of its energy is transfered to mode $\hat{\vc u}(0,4)$ 
through internal nonlinear interactions. Therefore, 
$|\hat{\vc u}(1,0)|<\lambda_e$, signals an upcoming extreme event. Since, in this quadrant,
we also have $\tilde D>D_e$, the indicator has correctly predicted the upcoming occurrence 
of an extreme event.

Quadrant III (false positives):
This quadrant corresponds to false positive predictions.
Since $|\hat{\vc u}(1,0)|<\lambda_e$, the indicator predicts an
upcoming extreme event. However, we have $\tilde D<D_e$ which 
implies no extreme events actually took place.

Quadrant IV (false negatives):
This quadrant corresponds to false negative predictions.
Since $|\hat{\vc u}(1,0)|>\lambda_e$, the indicator predicts no
upcoming extreme events. However, we have $\tilde D>D_e$ which 
implies that an extreme event actually took place.

Clearly, the quadrants III and IV are undesirable since 
the indicator incorrectly predicts the extremes or lack thereof.
However, only a small portion of the conditional density 
$p_{\tilde D|\lambda}$ resides in these quadrants, implying 
that $|\hat{\vc u}(1,0)|$ serves as a reliable indicator of extreme dissipation events.
In fact, the rate of false positive and false negative predictions are $0.85\%$ and $0.26\%$,
respectively; that is the overwhelming majority of extreme events are predicted correctly.

We also define the probability that an extreme dissipation event ($D>D_e$) takes place over the 
future time interval $[t+\tau_p,t+\tau_p+\Delta\tau_p]$ given $\lambda=|\hat{\vc u}(1,0,t)|$ at the current instant $t$. We denote this quantity by $P_{ee}$ and refer to it as the \emph{probability of upcoming extreme events}
which is defined by taking the marginal of the conditional probability $p_{\tilde D|\lambda}$, i.e., 
\begin{equation}
P_{ee}(\lambda) = \int_{D_e}^\infty p_{\tilde D|\lambda}(\zeta,\lambda)\id \zeta.
\end{equation}
For a given $\lambda=|\hat{\vc u}(1,0,t)|$, $P_{ee}(\lambda)$ measures the
probability that $D(s)>D_e$ for some time $s\in[t+\tau_p,t+\tau_p+\Delta\tau_p]$.

Figure~\ref{fig:R40_condPDF}(b) shows the probability of upcoming extreme dissipation events 
for the Kolmogorov flow. For relatively large values of $|\hat{\vc u}(1,0)|$ the probability of upcoming
extremes is virtually zero. As the mode $\hat{\vc u}(1,0)$ transfers its energy to the mode $\hat{\vc u}(0,4)$ 
and therefore $|\hat{\vc u}(1,0)|$ becomes relatively small,
the probability of upcoming extremes approaches one, signaling the high likelihood of an upcoming extreme event.
Below, we use the predictions obtained by $P_{ee}$ to decide whether or not to actuate the control.

\section{Controlled System}
Recall that the extreme energy dissipation events are instigated by
a nonlinear transfer of energy from mode $\hat{\vc u}(1,0)$ to mode $\hat{\vc u}(0,4)$. 
Therefore, it is natural to attempt to mitigate these extreme events by removing the excess energy 
from the mode $\hat{\vc u}(0,4)$. We accomplish this by designing the control term $\pmb\xi$ 
to have the form of a damping on mode $\hat{\vc u}(0,4)$. To this end, we set 
$\hat \xi_i(\vc k,t)\propto - \left[\hat u_i(\vc k_f,t)\delta_{\vc k,\vc k_f}+\overline{\hat u_i(\vc k_f,t)}\delta_{\vc k,-\vc k_f}\right]$
where $\vc k_f = (0,4)$. The complex conjugate term acting on the wave number $-\vc k_f$ is necessary to 
ensure that the resulting velocity field $\vc u(\vc x,t)$ is real valued. 
\begin{figure}
\centering
\includegraphics[width=.4\textwidth]{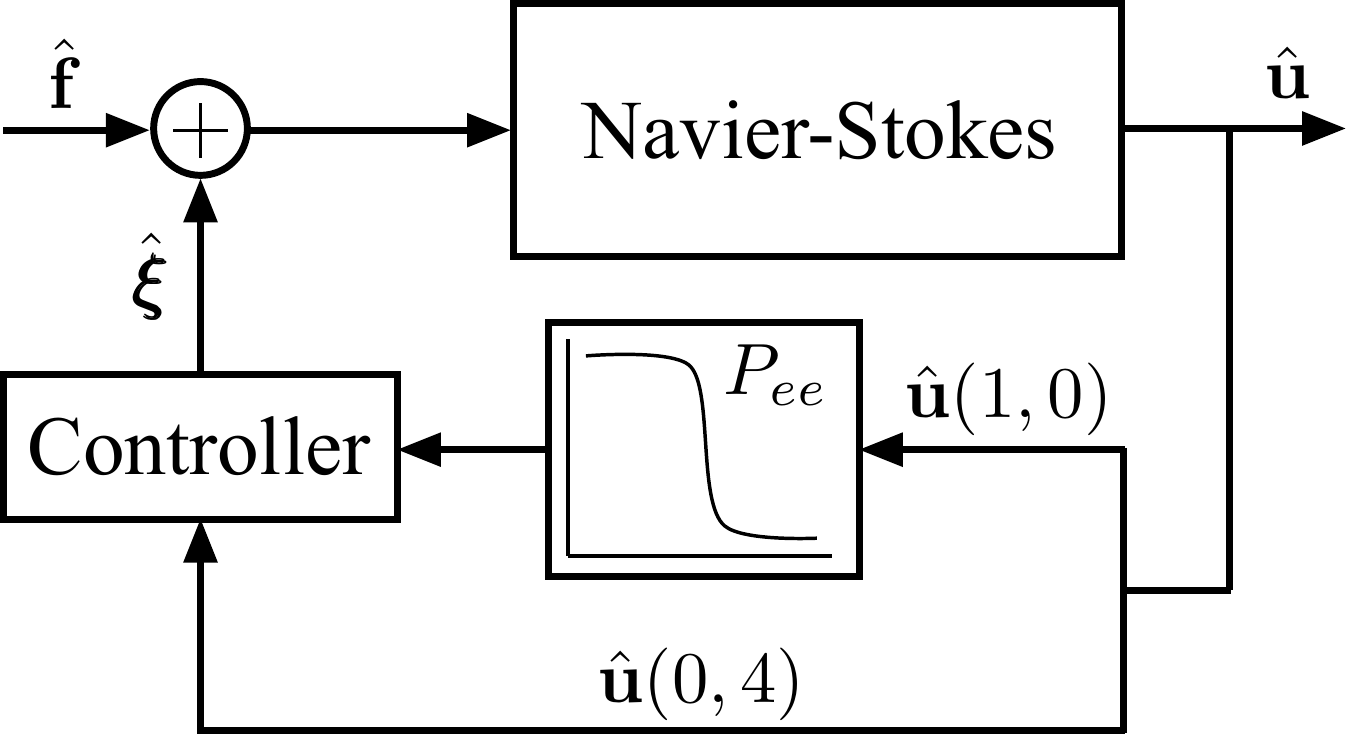}
\caption{The block diagram of the control strategy. The Fourier transforms of the external forcing, the velocity field and
the control term are denoted by $\hat{\vc f}$, $\hat{\vc u}$ and $\hat{\pmb \xi}$, respectively.
The Fourier mode $\hat{\vc u}(1,0)$ is used to measure the probability of upcoming extreme events $P_{ee}$.
The control is proportional to the Fourier mode $\hat{\vc u}(0,4)$ where coincidentally $\vc k_f=(0,4)$ is the 
wave number of the external forcing.}
\label{fig:block_diag}
\end{figure}

Note that this control only acts on the Fourier mode $\hat{\vc u}(\vc k_f)$
(and its complex conjugate counterpart $\hat{\vc u}(-\vc k_f)$).
Examining Eq.~\eqref{eq:nse} and neglecting the Navier--Stokes dynamics
for the moment, the controller acts on this mode
as $\partial_t\hat{\vc u}(\vc k_f,t)\propto -\hat{\vc u}(\vc k_f,t)$
which damps the excess energy content of the mode exponentially fast, 
$|\hat{\vc u}(\vc k_f,t)| \propto e^{-t}$.
\begin{figure*}
	\centering
	\includegraphics[width=.99\textwidth]{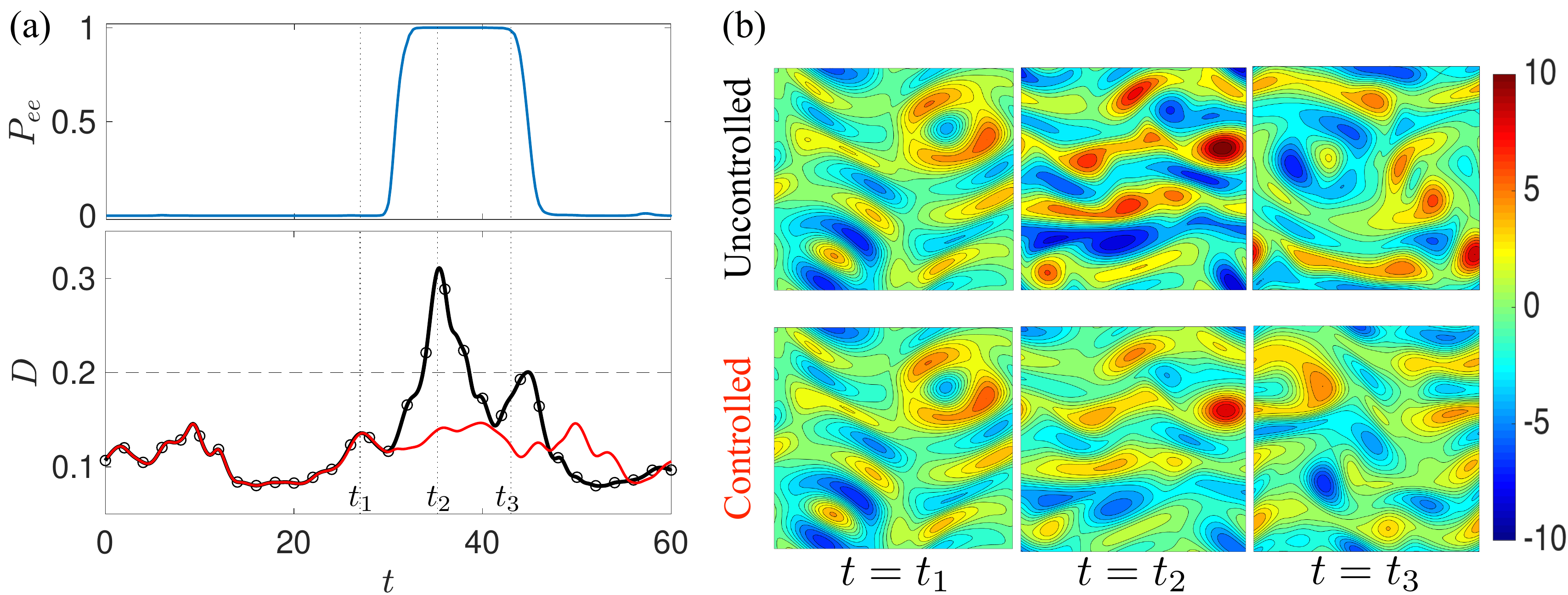}
	\caption{Controlled versus uncontrolled systems at $\re=40$.
		(a) Energy dissipation rate $D$ of the uncontrolled (black circles)
		and controlled (solid red) systems as a function of time.
		The horizontal dashed line marks the threshold $D_e=0.2$ for the extreme events.
		The top panel shows the probability $P_{ee}$ of upcoming extreme events. 
		(b) The corresponding vorticity fields at times 
		$t_1=27.0$, $t_2=35.2$ and $t_3=43.0$.}
	\label{fig:R40_vort}
\end{figure*}

We also would like the control to be actuated only when 
an extreme event is about to take place. To this end, we define
\begin{equation}
\hat \xi_i(\vc k,t)=-\frac{1}{\tau_c} P_{ee}(t) \left[\hat u_i(\vc k_f,t)\delta_{\vc k,\vc k_f}+\overline{\hat u_i(\vc k_f,t)}\delta_{\vc k,-\vc k_f}\right],
\end{equation}
where $P_{ee}(t)$ is shorthand for $P_{ee}(|\hat{\vc u}(1,0,t)|)$ (see figure~\ref{fig:R40_condPDF}(b)).
When the probability of upcoming extreme events is zero, the control is inactive since $P_{ee}=0$.
However, as that probability increases, the control term becomes active gradually until $P_{ee}$
approaches one and the controller becomes fully active. After the extreme event episode, the 
probability $P_{ee}$ decays back to zero and consequently the controller turns off. 
The parameter $\tau_c$ is the time lag between 
the control becoming fully active ($P_{ee}=1$) and the turbulent velocity field responding to the 
action of the control. Here, we set $\tau_c=\tau_p=1.0$. Figure~\ref{fig:block_diag}
summarizes the control strategy in a block diagram.

Taking the inverse Fourier transform, the control can be written in the 
physical space as 
\begin{equation}
\xi_i(\vc x,t) = -(2P_{ee}(t)/\tau_c)r_i(t) \cos (k_fx_2+\phi_i(t)),
\label{eq:cont_full}
\end{equation}
where $r_i$ and $\phi_i$ are the amplitude and phase of the Fourier mode $\hat{u}_i(\vc k_f,t)$,
respectively, so that $\hat{u}_i(\vc k_f) = r_ie^{\hat i\phi_i}$. Since the velocity field is divergence-free,
$\vc k_f\cdot \hat{\vc u}(\vc k_f)=0$, we have $r_2(t)=0$. As a result, $\xi_2(\vc x,t)\equiv 0$
and the control only acts on the horizontal component  $u_1(\vc x,t)$ of the velocity field.

Furthermore, numerical simulations suggest that, in the uncontrolled system, the phase $\phi_1$ oscillates 
around $-\pi/2$ with a small standard deviation. For instance, at $\re=40$, we have $\sigma(\phi_1)\simeq 0.03\pi$,
where $\sigma(\phi_1)$ denotes the standard deviation of $\phi_1$. 
As a result, the controller can be further simplified by assuming $\phi_1=-\pi/2$
which implies 
\begin{equation}
\xi_1(\vc x,t)=-(2P_{ee}(t)/\tau_c)r_1(t) \sin (k_fx_2),
\quad \xi_2(\vc x, t)=0.
\label{eq:cont_sin}
\end{equation} 
We note that this simplified control is 
a scalar multiple of the external forcing $\vc f$. The corresponding probability distributions
of the energy dissipation $D$ are nearly identical whether we use the full 
control~\eqref{eq:cont_full} or its simplified form~\eqref{eq:cont_sin}. 

Figure~\ref{fig:R40_vort} shows the closeup view of an extreme event
at $\re=40$; it compares the uncontrolled and controlled system
trajectories starting from the same initial condition.
Initially, the probability of upcoming extreme events 
is zero ($P_{ee}=0$) and therefore the control is inactive. As a result, 
the trajectories of the uncontrolled and controlled systems coincide. 
Around time $t\simeq 30$, the probability $P_{ee}$ increases towards one,
the control becomes active and the trajectory of the controlled system 
deviates from the uncontrolled system. Shortly after $t=30$, 
the uncontrolled system undergoes an extreme event ($D>D_e\simeq 0.2$).
However, the controlled system successfully evades any such event and its 
energy dissipation rate remain below the threshold $D_e$.
A longer time series of the energy dissipation of the controlled system is shown in figure~\ref{fig:D_ts}(b).

\begin{figure*}
	\centering
	\includegraphics[width=.99\textwidth]{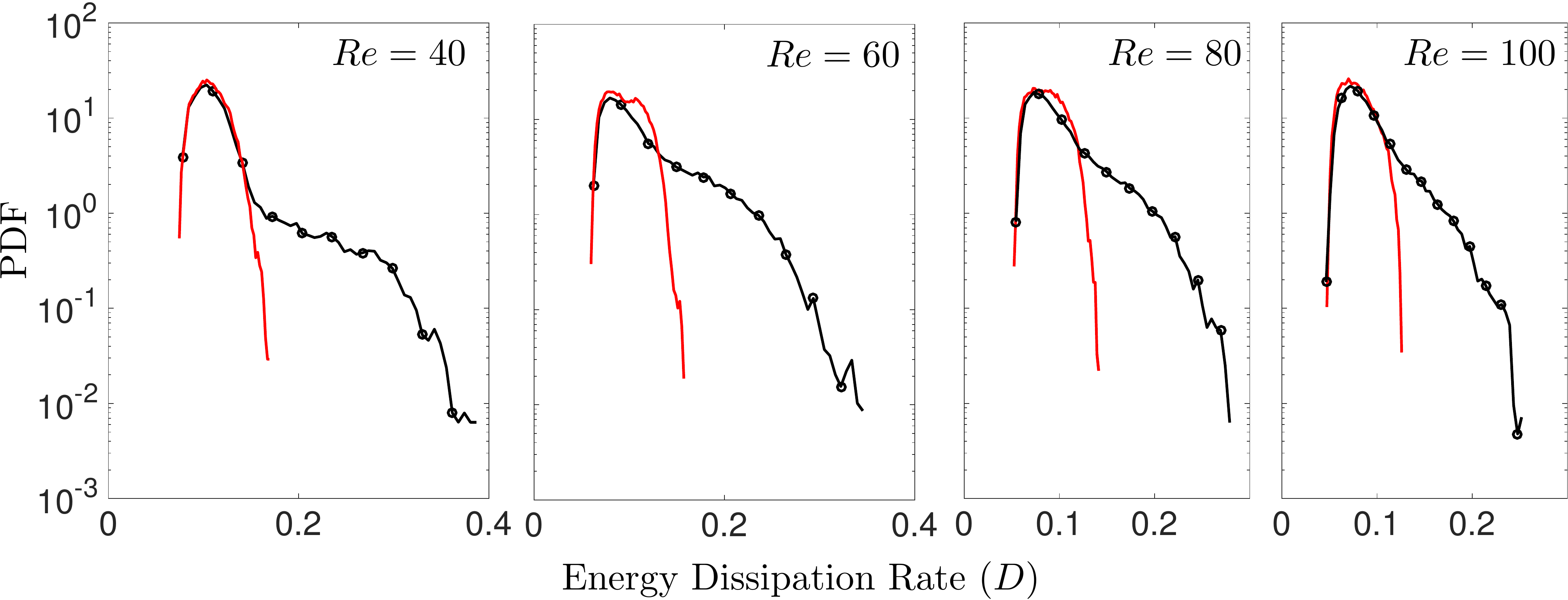}
	\caption{The probability density function (PDF) of the uncontrolled (black, circles) and controlled (solid red) systems
		at Reynolds numbers $\re=40$, $60$, $80$ and $100$. Each PDF is estimated from $50,000$ 
		data points.}
	\label{fig:D_pdf}
\end{figure*}

Figure~\ref{fig:D_pdf} shows the probability density function (PDF) of the energy dissipation
estimated from longterm simulations at Reynolds numbers $\re=40$, $60$, $80$ and $100$. 
The PDFs corresponding to the uncontrolled system
have heavy tails due to the extreme dissipation events. However, the PDFs of the controlled systems 
have no such heavy tails, indicating the successful mitigation of extreme events. 
Furthermore, the core of the PDFs (corresponding to non-extreme events) are very similar for both 
controlled and uncontrolled systems. This implies that the controller does not fundamentally change 
the nature of the flow; it only mitigates the extreme events, forcing the turbulent trajectories to 
stay on the core of the turbulent attractor (cf. figure~\ref{fig:schem_attractor}). 

Note that, while our controller acts on the mode $\hat{\vc u}(0,4)$, its activation 
is decided based on $P_{ee}$ which depends
on the precursor $|\hat{\vc u}(1,0)|$. We have also tried to actuate our controller based on 
the modulus of the controlled mode $|\hat{\vc u}(0,4)|$ instead of $|\hat{\vc u}(1,0)|$. 
While this control strategy suppresses some of the extreme events, it fails to remove the 
heavy tail events altogether.

We conclude by commenting on the possible 
experimental implementation of our control strategy. 
Kolmogorov-like flows have been studied in the laboratory experiments
by electromagnetically driving a thin layer of electrolyte~\cite{paret1997,burgess1999,Suri2017}. 
The electromagnetic force is exerted by an array of 
magnets with alternating magnetization, generating the 
sinusoidal forcing in equation~\eqref{eq:nse}. 

These lab experiments differ from our Kolmogorov flow in their boundary conditions
and that they are not strictly two-dimensional. As such, the mechanism underlying the extreme events
in the lab experiments should be investigated based on more accurate models, such as the
quasi-two-dimensional model developed in Ref.~\cite{suri2014}. 
If the extreme event mechanisms turn out to be similar to our Kolmogorov model, 
then our results are relevant to the experiments.

Since our control 
only acts on the same wave number as the external forcing, actuating 
the control in the lab experiments amounts to adjusting the 
magnitude of the external forcing (or equivalently the magnitude of 
the external magnetic field).
This magnitude would depend on the probability $P_{ee}$
which in turn depends on the magnitude of the Fourier mode $\hat{\vc u}(1,0)$.
Therefore, experimental implementation of our control strategy 
would require high-speed velocimetry~\cite{Westerweel2013,Poelma2016} so that the
control can be actuated in time to counteract the extreme events.
Since the precursor $|\hat{\vc u}(1,0)|$ corresponds to large scales 
(or equivalently the small wavenumber $\vc k=(1,0)$), only a low-pass 
filtered measurement of the velocity is sufficient.

\section{Conclusions}
A plethora of high-dimensional chaotic dynamical systems exhibit
spontaneous extreme events.
Recent advances in quantification and prediction of extreme events
show that only a few degrees of freedom might be directly involved in the 
formation of these events. This raises the prospect of designing 
low-dimensional controllers that only act on these few degrees of freedom 
in order to mitigate the extreme events. 

Here, we demonstrated the feasibility of such simple controllers on 
a turbulent fluid flow. While acting on a single Fourier mode, our controller
succeeded in suppressing all the extreme dissipation events. 

We emphasize two important features of our controller: low dimensionality and adaptivity. 
The low dimensionality of the controller is desirable as it facilitates its practical implementation. 
This is feasible due to the inherent low-dimensionality of the precursors to extreme events. 
Adaptivity refers to the fact that the controller is off most of the time,
and becomes active only when there is a probabilistic prediction of an imminent extreme event. 
Unlike classical methods for controlling chaos, our controller does not attempt to suppress 
chaos altogether. Instead, it only acts for relatively short periods of time
and exerts minimal interference with system dynamics. As such, the controlled system
is still chaotic but contains no extreme events. 

We propose that these two features (namely, low dimensionality and adaptivity) should 
form the basis of controlling extreme events more generally, beyond our fluid system.

\section*{Acknowledgments}
M.F. acknowledges fruitful conversations with Michael Schatz and Christopher Crowley (Georgia Tech). 
Figure~\ref{fig:schem_attractor} was generated using a publicly available code courtesy of Oleg Alexandrov.
This work has been supported through the ARO MURI W911NF-17-1-0306 and the ONR grant
N00014-15-1-2381. 

%\bibliographystyle{plain}
%\bibliography{bibliog.bib}
%apsrev4-2.bst 2019-01-14 (MD) hand-edited version of apsrev4-1.bst
%Control: key (0)
%Control: author (8) initials jnrlst
%Control: editor formatted (1) identically to author
%Control: production of article title (0) allowed
%Control: page (0) single
%Control: year (1) truncated
%Control: production of eprint (0) enabled
%

\end{document}